\documentclass[runningheads]{llncs}
\usepackage[T1]{fontenc}
\usepackage{graphicx}
\usepackage{amsmath}
\usepackage{siunitx}
\usepackage[caption=false]{subfig}
\usepackage{hyperref}
\usepackage{float}
\begin{document}

\title{A Feature-Rich Embedded NIDS with eBPF/XDP: Detector and Architecture Trade-offs}

\titlerunning{A Feature-Rich Embedded NIDS with eBPF/XDP}

%

\author{Shiqi Wu\inst{1,2}\orcidID{0009-0003-4677-7958}\thanks{S.~Wu and O.~Koshovyi contributed equally to this work and share first authorship.} \and
Oleksii Koshovyi\inst{1,2}\orcidID{0009-0009-2027-8652} \and
Georgios Pseiridis Pseiras\inst{2}\orcidID{0009-0001-3007-9923} \and
Victor Morel\inst{1}\orcidID{0000-0001-9482-8906} \and
Romaric Duvignau\inst{1}\orcidID{0000-0003-1268-9311}}
\authorrunning{S. Wu et al.}
\institute{Chalmers University of Technology, Gothenburg, Sweden\\
\email{shiqiw@student.chalmers.se}, \email{\{morelv,duvignau\}@chalmers.se}
\and
University of Gothenburg, Gothenburg, Sweden\\
\email{guskoshol@student.gu.se}
\and
Ericsson AB, Gothenburg, Sweden\\
\email{\{shiqi.wu,oleksii.koshovyi,georgios.pseiridis.pseiras\}@ericsson.com}}

%

%
%
%
\maketitle              
\begin{abstract}
Distributed Denial-of-Service (DDoS) attacks remain a serious threat to transport networks, with recent attack volumes exceeding 30 Tbps, and the telecommunications industry being the main target. Recent work has yet to study the impact of the hosting software architecture on network monitoring solutions, or to assess recent algorithms for improving attack detection. This paper presents a Network Intrusion Detection System (NIDS) for DDoS detection in transport networks, developed in collaboration with Ericsson. Building on a statistical baseline, we improve detection effectiveness with an Isolation Forest trained on a wider set of flow features, extracted by GoFlowMeter, our open-source Go implementation of CICFlowMeter, and we integrate eBPF/XDP so that the NIDS filters real traffic at the kernel level. We further compare three deployments, monolithic, Kafka-based, and gRPC-based microservices, on a Raspberry Pi 5 testbed replaying the CIC-DDoS2019 dataset as real network traffic. Detection quality is governed mainly by the choice of detector rather than by the transport: the Isolation Forest raises recall and F1 score (0.965 live in the monolithic variant) over the baseline by flagging low-volume attack windows that the baseline misses. The transport is not neutral, however: gRPC reaches almost the same accuracy as the monolithic variant while adding less than 2 milliseconds of transport time per window, whereas the asynchronous Kafka pipeline trails by roughly nine percentage points and adds about 27 milliseconds. These findings clarify the trade-off between detection quality and architectural overhead when deploying a NIDS on resource-constrained hardware.

\end{abstract}
%
%
%
\section{Introduction}
\label{sec:introduction}


Distributed Denial-of-Service (DDoS) attacks remain one of the most persistent threats to network infrastructure. Recent incidents have demonstrated attack volumes exceeding 30 Tbps, which is 10 times greater than 2 years before. Furthermore, the primary target of such attacks is telecommunications networks~\cite{cloudflare_aisuru_botnet_report,ferreira2026botnet}. As transport networks carry increasingly critical traffic, the need for effective and scalable intrusion detection at the network level has grown accordingly.

Existing Network Intrusion Detection Systems (NIDS) face two key challenges in this domain. First, detection approaches that rely on limited feature sets, such as packet counts alone~\cite{wickmannidschalmers}, struggle to distinguish sophisticated attack patterns from legitimate traffic surges. Second, the software architecture of the NIDS itself affects whether the system can meet the performance and maintainability demands of production transport environments; yet this architectural dimension is rarely studied along with the detection effectiveness.

Three developments motivate our approach, one per layer of the problem. At the mitigation layer, the Extended Berkeley Packet Filter (eBPF) with the Express Data Path (XDP) processes and drops malicious packets in the Linux kernel before memory is allocated for them, keeping the overhead of an attack minimal~\cite{xdp_10.1145/3281411.3281443}. At the detection layer, the Isolation Forest (IF)~\cite{isolation_forest_4781136} is a lightweight unsupervised algorithm with a short training time, whose detection performance can be improved further through fine-tuning and richer input features~\cite{wickmannidschalmers}. At the deployment layer, the choice between a monolithic architecture and microservices affects processing time and scalability, and Apache Kafka~\cite{apache_kafka,kafka_motivation1} and gRPC~\cite{grpc_motivation1} are the most widely used representatives of asynchronous and synchronous inter-service communication; we therefore evaluate both.

\subsection{Research questions}
\label{sec:research_questions}

We defined two research questions (RQs) to guide our research:

\begin{enumerate}
\item How can we improve DDoS detection effectiveness in transport networks by proposing an improved detection approach that combines kernel-level eBPF filtering with adaptive machine learning?
\item How do different software architecture patterns (monolithic, Kafka-based microservices, and gRPC-based microservices) impact the software performance of NIDS when integrated with different detection algorithms?
\end{enumerate}

To answer these questions, this paper proposes a NIDS that combines eBPF/XDP kernel-level filtering with an adaptive Isolation Forest trained on a rich flow-level feature set extracted by our open-source tool GoFlowMeter~\cite{goflowmeter}, and empirically compares three deployment architectures (monolithic, Kafka-based, and gRPC-based microservices) on the CIC-DDoS2019 dataset~\cite{cicddos2019_paper} using a Raspberry Pi 5 testbed.

This work was carried out in collaboration with Ericsson's Radio and Transport Engineering division, whose transport products form the backbone of telecommunication networks, the primary target of recent large-scale DDoS attacks.


\subsection{Contribution}
\label{sec:contribution}
This paper makes the following contributions.

\begin{itemize}

\item We integrated eBPF/XDP packet filtering into the NIDS, so that it captures, processes, and filters real traffic on a network interface while protecting device resources.


\item We implemented and compared three deployment architectures (monolithic, Kafka-based, and gRPC-based microservices) and quantified their impact on detection quality, latency, and memory, showing that the gRPC variant incurs an order of magnitude less transport overhead than the Kafka variant.


\item We implemented GoFlowMeter~\cite{goflowmeter}, an open-source Golang re-implementation of CICFlowMeter~\cite{cicflowmeter1}, and fine-tuned the Isolation Forest on the extracted 80-feature flow vectors, addressing the single-feature limitation of the previous system~\cite{wickmannidschalmers}.


\item We implemented a standalone sequential replayer that replays more than 100 GB of CIC-DDoS2019~\cite{cicddos2019_paper} traffic against the NIDS over a real network interface.



\end{itemize}

\subsection{Paper outline}
The remainder of this paper is organised as follows. Section~\ref{chapter:background} introduces the background and Section~\ref{chapter:related_work} reviews related work. Section~\ref{chapter:methods} presents the system and the experimental setup, Section~\ref{chapter:results} reports the results, Section~\ref{chapter:discussion} discusses them, and Section~\ref{chapter:conclusion} concludes.





\section{Background}

\label{chapter:background}

This section summarises the concepts the paper builds on: DDoS attacks and the dataset used for evaluation, network intrusion detection based on flow features, the Isolation Forest detector, kernel-level packet filtering with eBPF/XDP, and the two microservice communication frameworks compared in our experiments.

\subsection{DDoS Attacks and the CIC-DDoS2019 Dataset}
\label{sec:ddos}

A DDoS attack floods a victim with traffic from many sources to exhaust its resources and deny service to legitimate users. Sharafaldin et al.~\cite{cicddos2019_paper} classify modern DDoS attacks by execution mechanism into reflection-based attacks, in which spoofed requests cause third-party servers to overwhelm the victim with responses, and exploitation-based attacks, such as SYN and UDP floods, which exhaust the victim's resources directly.

We evaluate on CIC-DDoS2019~\cite{cicddos2019_paper}, the most recent publicly available dataset focused exclusively on DDoS attacks. It provides two days of labelled traffic covering both reflection-based and exploitation-based attacks, with benign background traffic generated from the B-profile of 25 simulated users~\cite{CIC-IDS2017}, and ships with an 80-feature flow analysis extracted by CICFlowMeter. Older public datasets such as CAIDA2007~\cite{caida2007} and CIC-IDS2017~\cite{CIC-IDS2017} were discarded as outdated or not DDoS-focused~\cite{Goldschmidt_2025}. A limitation is that CIC-DDoS2019 traffic is synthetically generated; realistic captured traffic is typically private and unlabelled.

\subsection{Intrusion Detection Systems}
\label{sec:ids}

Intrusion detection systems are commonly categorised as host-based (HIDS) or network-based (NIDS)~\cite{hids_vs_nids}. A NIDS inspects traffic at network vantage points and is the natural fit for DDoS defence: the attack signal is distributed across aggregated traffic rather than visible on a single host, detection can happen upstream before targets saturate, and it connects directly to mitigation at the ingress.

\subsubsection{Network Flows and Sliding Time Windows}
\label{sec:flows_windows}

A network flow is a sequence of packets sharing the same 5-tuple (source and destination IP address, source and destination port, and protocol); grouping both directions yields a bidirectional flow. Flow-level statistics summarise connection behaviour more compactly than individual packets and are the standard input for flow-based intrusion detection~\cite{cicflowmeter1,cicflowmeter2}. Because a traffic stream has no natural boundaries, packets are additionally grouped into sliding time windows of fixed length, and each completed window is classified independently; a flow that spans several windows contributes only a sub-flow to each.

\subsubsection{CICFlowMeter}
\label{sec:cicflowmeter_background}

CICFlowMeter~\cite{cicflowmeter1,cicflowmeter2} is an open-source tool from the Canadian Institute for Cybersecurity that assembles packets into bidirectional flows and extracts 80 statistical features (e.g., flow duration, packet counts, inter-arrival times). Its feature set is the de facto standard in flow-based intrusion detection research and defines the feature space used throughout this paper.

\subsection{Isolation Forest}
\label{sec:machine_learning}

The Isolation Forest~\cite{isolation_forest_4781136,wickmannidschalmers} is an unsupervised machine learning algorithm designed for anomaly detection. Introduced by Liu et al.~\cite{isolation_forest_4781136} in 2008, this approach constructs an ensemble of isolation trees from the data. Each isolation tree is built by recursively partitioning the dataset using randomly selected features and randomly chosen split values within their respective ranges, until each data point is fully isolated in its own leaf node. The anomaly score of a point is derived from its average path length across all trees in the forest: the shorter this average path length, the higher the likelihood that the point is an anomaly. Since anomalous points are rare and require fewer partitions to be isolated, their average path length is shorter.

Although supervised models such as LSTM achieve the highest reported detection scores~\cite{xbpf_xdp_smartx,cicddos2019_paper,tebbaa2025mitigating}, IF is unsupervised: it needs no labelled training data, is lightweight~\cite{perfromance_impact}, and handles zero-day attacks that are absent from training data better~\cite{wickmannidschalmers}, which suits the operational setting targeted here.

\subsection{Fast filtering with eBPF/XDP}
\label{sec:ebpf_xdp}
The Extended Berkeley Packet Filter (eBPF) is a Linux kernel virtual machine that allows user-supplied programs to run on kernel-level events, exchanging data with user space through shared key/value maps~\cite{xdp_10.1145/3281411.3281443}. The Express Data Path (XDP) hooks such programs directly into the network driver, so packets can be dropped, passed, or redirected before the kernel allocates memory for them, at rates exceeding 20 million packets per second~\cite{xdp_10.1145/3281411.3281443,xbpf_xdp_smartx}. This makes eBPF/XDP well suited to shedding DDoS traffic at minimal cost.

\subsection{Microservice Communication Frameworks}
\label{sec:microservice_frameworks}

A microservice architecture decomposes a system into independently deployable services, and the inter-service communication style directly affects performance. We compare the two most common styles through widely used representatives: asynchronous messaging via a broker (Apache Kafka) and synchronous remote procedure calls (gRPC).

\subsubsection{Apache Kafka}

Apache Kafka~\cite{apache_kafka} is a distributed publish-subscribe broker: producers write messages to topics, and consumers read them at their own pace. This decoupling provides buffering (bursts are absorbed on disk rather than by the consumer), horizontal scaling (consumers in a group share topic partitions), and persistence (messages survive consumer restarts). The cost is that every message pays for serialisation, broker round-trips, and the consumer fetch cycle, so latency depends on broker configuration rather than on function-call overhead.

\subsubsection{gRPC}

gRPC~\cite{grpc_website} is a remote procedure call framework using Protocol Buffers~\cite{protocol_buffer} as its interface definition and compact serialisation format. It follows a synchronous request-response model: the client blocks until the server responds, so per-request latency includes the round trip plus (de)serialisation, but avoids broker persistence and buffering and gives the caller its result within the same call.

\section{Related Work}

\label{chapter:related_work}

This section reviews existing work on NIDS for DDoS detection and prevention, covering statistical and machine learning detection approaches as well as kernel-level packet filtering with eBPF/XDP.

\subsection{NIDS for DDoS Defence}

Research on NIDS for DDoS attacks ranges from statistical models to machine learning techniques. This section first reviews the previous work at Ericsson by Wickman and Rygaard~\cite{wickmannidschalmers}, which established a statistical baseline and explored unsupervised anomaly detection using Isolation Forest (IF). It then covers supervised approaches based on Convolutional Neural Networks (CNN), Long Short-Term Memory (LSTM) networks, and Random Forests (RF), and finally examines recent unsupervised methods that address the reliance on labelled datasets.

Wickman and Rygaard~\cite{wickmannidschalmers}, in a previous master's thesis conducted at Ericsson, combined machine learning techniques with probabilistic data structures, namely the Count-Min Sketch and the HeavyKeeper algorithm. A data sketch is a compact probabilistic structure that summarises a high-volume traffic stream using a small and fixed amount of memory, at the cost of small and bounded estimation errors. The Count-Min Sketch provides approximate per-source packet frequency counts, while HeavyKeeper identifies the heavy hitters, that is the small number of sources responsible for a disproportionate share of the traffic. IF was chosen because, as an unsupervised approach, it requires no labelled training data, which not all of their datasets provided. The authors developed two IF detectors, one for the whole traffic flow (packet counts per time window) and one for individual IPs. During a warmup procedure, threshold IFs are trained on benign traffic, with the threshold refined after each time window. After warmup, the system filters out malicious traffic that exceeds the threshold. Evaluated on CAIDA2007~\cite{caida2007} and Ericsson's internal datasets, their IF used only one feature, the packet count, and reached a detection accuracy of 73.63\%, below the 96.09\% of the statistical approach. We are improving the IF approach by extracting and using more features to identify DDoS attacks. Their statistical method will be our baseline in this study, and ``Baseline'' refers to it in the remainder of this paper.

CNNs perform classification based on labelled training data and are well suited to two-dimensional data such as images. Shaaban et al.~\cite{nids_cnn_Shaaban} restructured one-dimensional network feature vectors into two-dimensional matrices with padding and achieved 99~\% accuracy on a simulated network dataset and the NSL-KDD dataset. Akgun et al.~\cite{nids_cnn} developed an inception-like CNN model, emphasising data preprocessing by removing redundant records and applying feature selection to the CIC-DDoS2019 dataset, and achieved 99.99~\% binary and 99.30~\% multiclass classification accuracy. However, CNN architectures are inherently designed for spatial data, while network traffic is sequential, and converting one-dimensional packet features into two-dimensional matrices introduces an artificial structure that may neglect important temporal patterns.

To utilise these temporal patterns, Shurman et al.~\cite{nids_lstm} developed an LSTM-based model to detect Distributed Reflection Denial-of-Service attacks, achieving a test accuracy of 99.19~\% on the CIC-DDoS2019 dataset with a three-layer, 128-unit architecture. Djama et al.~\cite{nids_lstm_djama} proposed a hybrid architecture in which convolutional layers extract spatial features and recurrent layers capture temporal dependencies, reaching an accuracy of 99.96~\% on the CICIDS2017 dataset.

RF is an ensemble learning method that constructs multiple decision trees to improve accuracy and reduce overfitting. Bhawsar et al.~\cite{nids_random_forest_bhawsar} compared RF and Support Vector Machines (SVM) for DDoS detection on the CIC-DDoS2019 dataset. Their RF model achieved an accuracy of 99.92\% and produced fewer false negatives than the SVM model, but its execution time was around ten times larger, which could affect performance in time-sensitive use scenarios.

CNN, LSTM, and RF are all supervised learning approaches that require labelled datasets, and the quality of the dataset affects the detection results. Moreover, a pre-trained model might become outdated due to the rapid growth of the approaches, characteristics, and amounts of DDoS attacks, which makes unsupervised learning potentially helpful. Vinothina et al.~\cite{vinothina2025} proposed an anomaly detection scheme that combines an autoencoder with XGBoost for feature selection. The autoencoder is trained exclusively on normal traffic and flags data that exceeds a reconstruction error threshold, which reduces data dimensionality and computational overhead on the UNSW-NB15 dataset. Ghani et al.~\cite{ghani2025} applied the IF algorithm to detect anomalies such as excessive login attempts and abnormal port activity. The algorithm isolates outliers by randomly partitioning data, identifying new attack patterns without prior labelling, and their system uses this unsupervised step as a primary warning mechanism against threats like backdoor exploits and DDoS attacks before passing the data to a supervised classifier. These approaches show that unsupervised learning can model normal traffic patterns and detect deviations caused by unknown attack vectors.

\subsection{Fast Packet Processing with eBPF/XDP}
Farasat~et~al.~\cite{xbpf_xdp_smartx} proposed a framework for NIDS with an eBPF/XDP filtering procedure. The researchers used eBPF/XDP to process and pass/drop packets. They also compared different ML algorithms, including SVM, Decision Tree, LSTM and Bidirectional LSTM. The detection and prevention performance of the eBPF/XDP framework with the ML algorithm were examined using a generated dataset~\cite{dataset-xdp-farasat2024} and the Bidirectional LSTM (BiLSTM) algorithm showed the best result, with an accuracy of 99.3\%, an F1 score of 99.3\%, and a ROC AUC of 99.9\%. The authors used CICFlowMeter-V4.0 to extract features from the traffic flow and SelectKBest to select the top 20 features.

\subsection{Gap Analysis}

A review of existing NIDS literature indicates a trend of reliance on supervised machine learning models for DDoS detection, while the application of unsupervised learning methods could be investigated further. Both supervised models, such as CNN and RF, and unsupervised models, like Autoencoders, have demonstrated high detection rates when evaluated on standard datasets such as CIC-DDoS2019. However, for supervised models, securing accurately labelled network traffic for training is difficult and time-consuming in practical environments, including the specific operational context at Ericsson Radio and Transport Engineering. Network configurations change frequently, and new attack vectors emerge constantly, making a reliance on static labelled data impractical. Consequently, a gap exists for an unsupervised method that leverages multiple network features to identify anomalies without requiring labelled training data.

Beyond the challenge of data labelling, bridging this research and industry gap requires focusing on operational performance, a factor often neglected in existing studies. Current research primarily focuses on maximising detection accuracy, yet various models and methods require relatively significant computational resources. For instance, Bhawsar et al.~\cite{nids_random_forest_bhawsar} demonstrated that RF execution times are longer than those of simpler models, and neural networks introduce inherent latency during feature extraction and classification. In practical network environments, an NIDS must process packets at high speeds to mitigate DDoS attacks before they overwhelm the infrastructure. Therefore, optimising system performance parameters, such as memory usage and latency, is a critical requirement for transferring the effective unsupervised detection methods from academia to the industry.

\section{Methods}

\label{chapter:methods}

This section presents the proposed NIDS and its evaluation components: a standalone traffic replayer, the GoFlowMeter feature extractor, the window-level ground-truth derivation, an offline hyperparameter probe, three deployment architectures, the eBPF/XDP integration, and the Raspberry Pi 5 testbed on which all experiments run.

\subsection{Traffic Replay and Feature Extraction}

To design and implement a NIDS that meets the operational constraints of a network infrastructure, such as limited memory, bounded latency, and continuous processing under high traffic volume, practicality in terms of performance impact is an important aspect to consider. To replay the network traffic in the dataset in an efficient yet continuous way, a sequential replayer is proposed in Section~\ref{sec:sequential_replayer}. It aims to reduce the memory pressure of loading the large dataset, such as CIC-DDoS2019~\cite{cicddos2019_paper}. 

\subsubsection{Sequential Replayer}
\label{sec:sequential_replayer}
To evaluate detection performance and system impact, we replay CIC-DDoS2019~\cite{cicddos2019_paper}, a dataset of over 150 GB of traffic split into 1000 PCAP files, from a standalone attacker program to the NIDS device over a real network interface. The replayer preloads the next PCAP file while the current one is being sent, so replay proceeds without pauses, and it preserves the original inter-packet delays by waiting whenever the elapsed replay time runs ahead of the capture timestamps~\cite{wickmannidschalmers}. Before transmission, a warmup filter forwards only known-benign traffic during a configurable startup period so that the NIDS can stabilise, and a frame rewriter redirects each Ethernet frame to the victim interface while keeping IP addresses and ports unchanged; frames are sent through a raw socket. The maximum replay rate is quantified in Section~\ref{sec:limitation_replayer}.


\subsubsection{Network Feature Analysis Using GoFlowMeter}
\label{sec:GoFlowMeter}

CICFlowMeter (Section~\ref{sec:cicflowmeter_background}) is implemented in Java and cannot be embedded naturally into our Go-based NIDS. We therefore implemented GoFlowMeter~\cite{goflowmeter}, an open-source native Go library that assembles the packets of each time window into bidirectional flows and extracts the same 80 statistical features as CICFlowMeter, so its output remains fully compatible with existing datasets and models.

GoFlowMeter produces one feature vector per bidirectional flow, whereas the detector consumes a single vector per time window, so the engine aggregates the per-flow vectors into one window-level vector. For each completed window, the packets buffered during that window are passed to GoFlowMeter, which assembles them into bidirectional flows and computes the 80 CICFlowMeter-compatible features (Section~\ref{sec:cicflowmeter_background}) per flow. Each flow $f$ is mapped to an 80-dimensional vector $x_f$ in a fixed feature order, and the window vector $\bar{x}$ is the element-wise arithmetic mean over the set $F$ of flows observed in the window:
\begin{equation}
\bar{x}_k = \frac{1}{|F|}\sum_{f \in F} x_{f,k}, \qquad k = 1,\dots,80 .
\label{eq:window_aggregation}
\end{equation}
Any NaN or infinite component (for example, a rate computed over a zero-duration flow) is replaced by zero, and a window that contains no flows is represented by the 80-dimensional zero vector. The per-IP feature vectors used by the IP-level detector are built with the same element-wise mean, restricted to the flows whose source address is that IP.

\subsubsection{Ground Truth Determination}
\label{sec:ground_truth_determination}

The CIC-DDoS2019 dataset provides pre-classified network flow labels alongside flow features extracted by CICFlowMeter. However, these labels are coarse-grained, as each label covers the entire duration of a given flow. In this paper, incoming network packets are buffered and grouped into sliding time windows. Consequently, a flow within a single window may represent only a sub-flow of the full flow described by the dataset label. Since the objective is to identify time windows containing malicious traffic, a mechanism is required to derive window-level ground truth from the flow-level labels provided by the CIC-DDoS2019 dataset.

\begin{figure}[htb]
  \centering
  \includegraphics[width=0.4\linewidth]{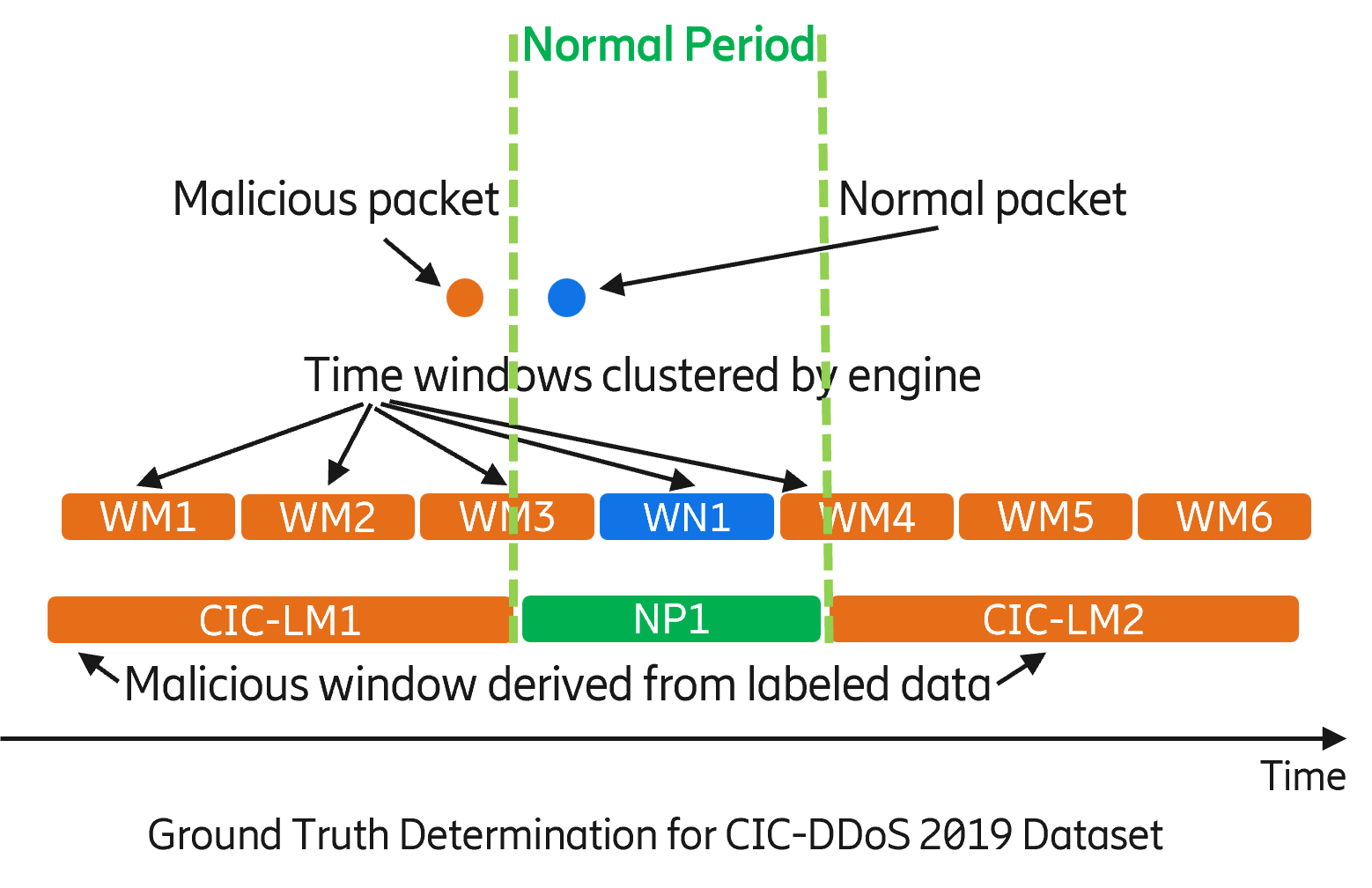}
  \caption[Ground truth determination method for incoming packets and sliding time windows]{Ground-truth determination for packets and sliding windows. Orange and blue dots mark packets inside and outside a labelled malicious interval (CIC-LM), respectively. Windows WM1--WM6 overlap a malicious interval and are labelled malicious, while WN1 lies entirely within a normal period (NP1) and is labelled normal.}
  \label{fig:ground_truth}
\end{figure}

As illustrated in Figure~\ref{fig:ground_truth}, per-flow labels are merged into malicious time intervals. There could be multiple flows in one time duration, and that duration is classified as malicious if any of the flows inside that duration is classified as malicious. Next, each incoming network packet is classified as malicious if it falls within a malicious interval, and as normal otherwise. A sliding window is then classified as malicious if it overlaps with any malicious interval; it is considered normal only if the entire window falls within a normal period.

Packet and window labels are derived independently, each solely by overlap with the malicious intervals. The rule is conservative in the security-relevant direction: any window in which an attack is in progress, even partially, is labelled malicious, matching the operational requirement that the detector raise an alert as soon as malicious activity is present, while a window counts as normal only if no part of it overlaps a malicious interval.

\subsection{Fine-Tuning of Isolation Forest}

\label{sec:probe}

To select Isolation Forest hyperparameters we built an offline probe, separate from the live engine, that parses every PCAP file once, groups packets into one-second windows under the same warmup stage as the engine, and caches the per-window packet count together with the aggregated feature vector of Equation~\ref{eq:window_aggregation}.

The probe varies the three standard IF hyperparameters: \texttt{num\_trees} (ensemble size), \texttt{sample\_size} (subsample used to build each tree), and \texttt{anomaly\_fraction} (the expected anomaly proportion, which sets the decision threshold).

It sweeps a grid in which \texttt{num\_trees} takes the values 50, 100, and 200, \texttt{sample\_size} the values 128, 256, and 512, and \texttt{anomaly\_fraction} the values 0.01, 0.05, 0.1, and 0.2, in two input modes: the per-window packet count alone, and the full 80-dimensional vector. The top-ranked combination is then used in the live monolithic, Kafka-based, and gRPC-based deployments.

\subsection{Software Architecture}

All three configurations run the same detection pipeline and differ only in how the engine and the detector communicate.

\subsubsection{Monolithic Architecture}

Monolithic software architecture usually refers to software that integrates the entire application into a single process~\cite{monolithic_definition}. In this paper, the monolithic NIDS is implemented as one Go application where packet ingestion, aggregation, detection, mitigation, logging, and visualisation are executed in the same operating system process. This design is used as a baseline architecture because it minimises orchestration overhead and allows direct in-memory communication between components.

\begin{figure}[htbp]
    \centering
    \subfloat[Monolithic]{
        \includegraphics[width=0.8\linewidth]{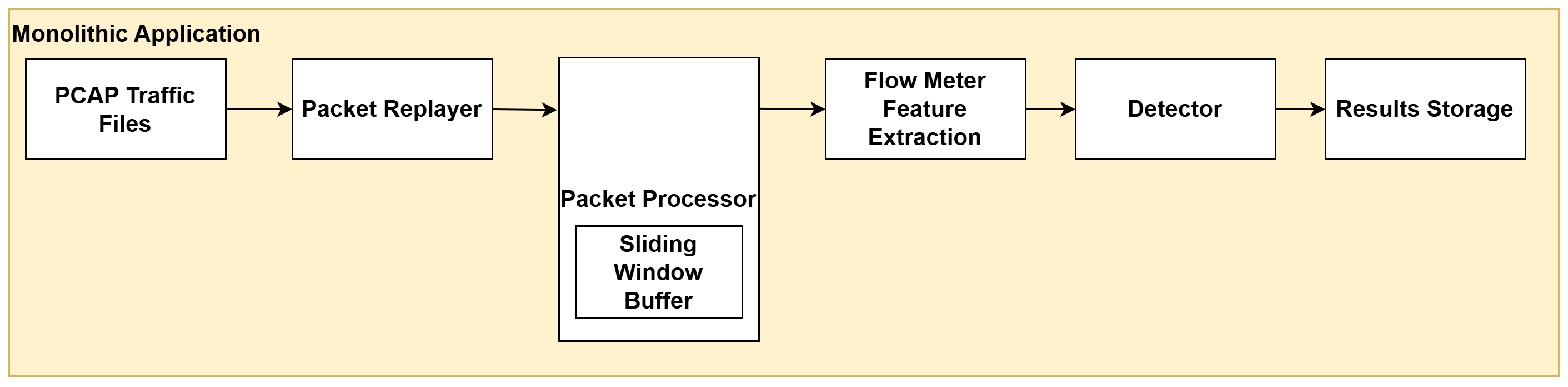}}
    \hfill
    \subfloat[Kafka-based]{
        \includegraphics[width=0.99\linewidth]{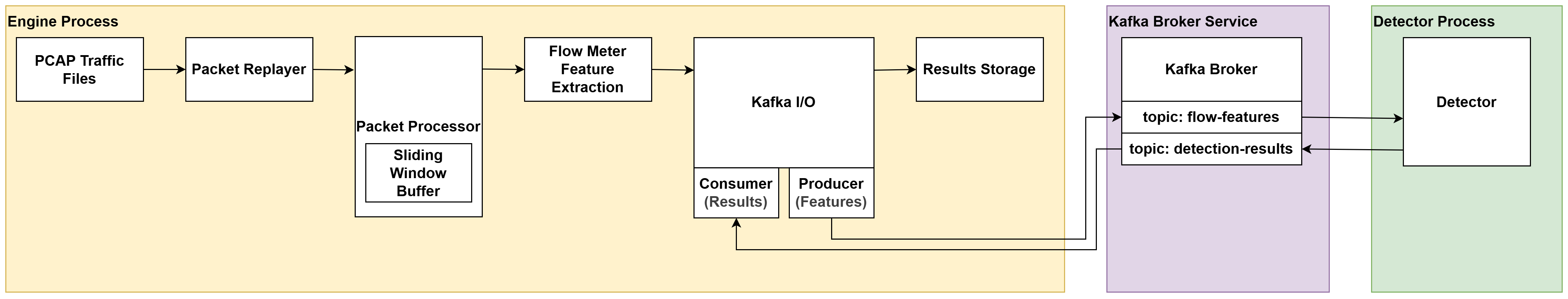}}
    \hfill
    \subfloat[gRPC-based]{
        \includegraphics[width=0.99\linewidth]{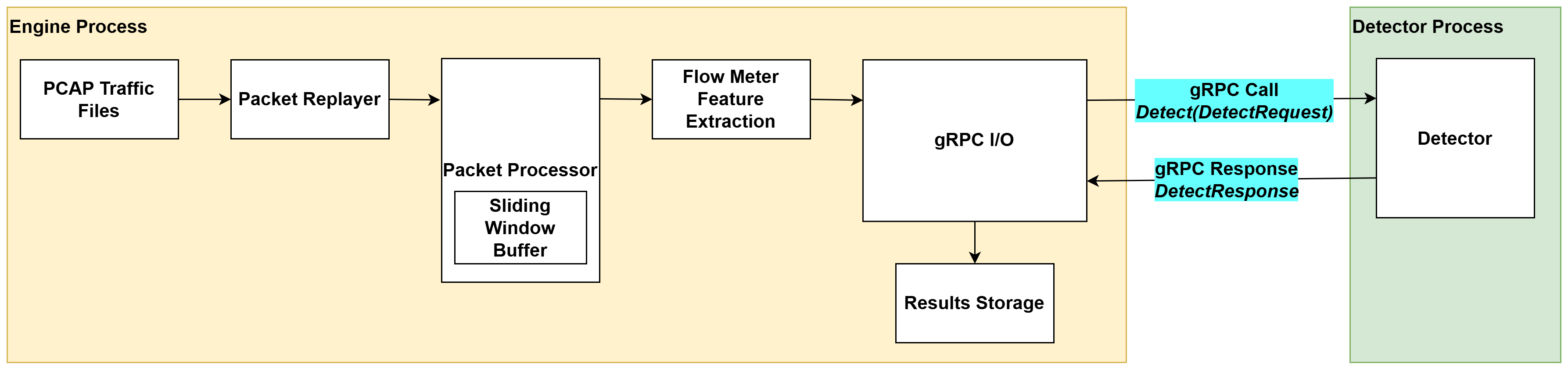}}
    \caption{The three evaluated NIDS architectures. The detection pipeline is identical; only the engine-to-detector transport differs.}
    \label{fig:architectures}
\end{figure}

The engine coordinates traffic capture, sliding-window management, detection, online model updates, and output generation in one process, so features extracted by GoFlowMeter are consumed directly in memory without serialisation. Each incoming packet updates per-window and per-source statistics and, when mitigation is enabled, is checked against the current per-IP rate limit before entering the active windows. Every completed window is classified at two levels: the window level decides whether aggregate behaviour indicates an attack, and the per-IP level identifies suspicious sources. The detector runs either in baseline mode, an IQR-based threshold strategy, or in machine-learning mode using Isolation Forest with count-based or feature-based input, selected through configuration. After each decision the thresholds are propagated to the rate limiter and the model is updated online with non-malicious observations, so it adapts to evolving normal traffic; during detected attacks the per-IP threshold is frozen to avoid instability. Figure~\ref{fig:architectures}a shows the pipeline.

\subsubsection{Microservice Architectures}

This study uses two microservice architectures that differ only in the communication framework placed between the NIDS engine and the detector. One uses Apache Kafka and the other uses gRPC. The general characteristics of these two frameworks are described in Section~\ref{sec:microservice_frameworks}; this section describes how each of them is integrated into the detection pipeline.

\paragraph{Kafka-based Architecture}

The Kafka variant places a broker between feature extraction and detection. The engine serialises each completed window into a feature message and produces it to a topic; a separate detector process consumes the message, runs the configured detection mode, and produces the outcome to a result topic that the engine consumes asynchronously. Everything downstream of the transport is identical to the monolithic version, and the buffering, scaling, and persistence properties of Section~\ref{sec:microservice_frameworks} apply directly (Figure~\ref{fig:architectures}b).

\paragraph{gRPC-based Architecture}

The gRPC variant replaces the in-memory call with a unary RPC: the engine is the client and the detector a stand-alone server exposing a single \texttt{Detect} method. The request carries the window identifier, the aggregated window feature vector, the per-IP feature vectors and packet counts, and the detection mode flags; the response returns the attack decision, anomaly score, threshold, malicious sources, and the time spent inside the detector. Because the call is synchronous, each result is available in the same call that produced it (Figure~\ref{fig:architectures}c). All three variants share the same detector code, which keeps the detection logic comparable.

\subsection{eBPF/XDP}

\begin{figure}[H]
    \centering
    \includegraphics[width=0.63\linewidth]{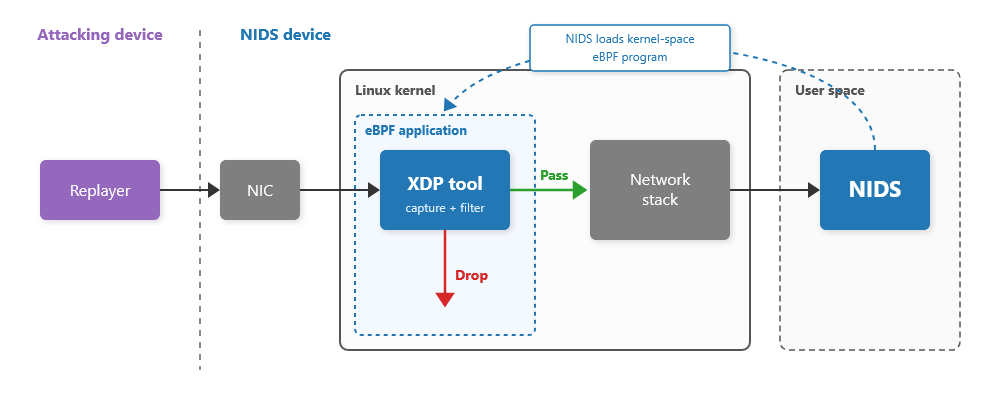}
    \caption[NIDS with eBPF integration]{NIDS with eBPF integration}
    \label{fig:nids_ebpf}
\end{figure}

Unlike the previous NIDS~\cite{wickmannidschalmers}, in which replay and processing ran inside one program and no traffic crossed a network interface, the present system captures and filters real traffic through eBPF/XDP. The rate limiter of the previous system is kept as the indicator of malicious sources~\cite{xbpf_xdp_smartx}: any source IP whose traffic exceeds the detector-derived threshold is added to a blocklist and dropped in subsequent windows (Figure~\ref{fig:nids_ebpf}).

\subsubsection{Kernel Integration and Filtering Logic}
The eBPF program is written in C, compiled, and loaded into the kernel by the engine, which attaches its XDP hook to the monitored interface. For each arriving packet the hook looks up the source address in a shared BPF blocklist map: a hit triggers XDP\_DROP, discarding the packet before the kernel allocates memory for it, while a miss triggers XDP\_PASS and forwards the packet metadata (addresses, ports, protocol, TCP flags, direction, and length) to user space through a perf ring buffer. On the Go side, a processor component reads these events and feeds them to the detection pipeline, and an enforcer component writes malicious addresses into the blocklist map. At the end of each window the Isolation Forest computes the rate threshold; sources exceeding it are confirmed as attackers and added to the map, after which the kernel drops their traffic automatically.

\subsection{Experimental Setup and Evaluation}

All experiments run on a Raspberry Pi 5 with a 2.4 GHz quad-core Arm Cortex-A76 CPU and 16 GB of RAM~\cite{raspberry_pi_spec}, running 64-bit Linux. The constrained platform exposes the cost of each architectural choice more clearly than a workstation would, approximates deployment near the network edge, and, being fixed and dedicated across all experiments, removes hardware variability from the comparison.

All processes of a configuration run on the same board: the monolithic engine as a single process, the Kafka engine, broker, and detector as three processes, and the gRPC engine and detector as two. Inter-process communication uses the loopback interface, so the comparison isolates the overhead characteristic of each communication framework rather than external network conditions.

\subsubsection{Evaluation Metrics}
\label{sec:evaluation_metrics}

Detection quality is measured with the standard accuracy, precision, recall, and F1 score, computed over the window-level ground truth of Section~\ref{sec:ground_truth_determination}.


The software performance metrics consist of three time measurements and the memory consumption of the engine process. The three time measurements are processing time, detector time, and transport time, and their precise definitions depend on the software architecture. They are introduced together with the corresponding equations in Section~\ref{sec:software_performance_results}.

\subsubsection{Overview of Experiments}

Three experiments are reported in the next section. The first tunes the Isolation Forest hyperparameters with the offline probe of Section~\ref{sec:probe} and fixes the configuration that the live system uses. The second compares detection quality, using the metrics defined above, between the statistical baseline and Isolation Forest across the monolithic, Kafka-based, and gRPC-based architectures. The third compares software performance, namely processing, detector, and transport time, together with memory consumption, across the same six architecture--detector combinations.

\section{Results}
\label{chapter:results}
This section reports three experiments: the Isolation Forest hyperparameter search, the detection quality comparison between the baseline and the Isolation Forest across the three architectures, and the software performance comparison in time and memory.

\subsection{Hyperparameter Tuning Results}

The probe of Section~\ref{sec:probe} swept all 72 configurations (36 per input mode) and ranked them by F1 score and accuracy over three runs. Table~\ref{tab:if_probe_input_modes} summarises the outcome. The anomaly fraction dominates: every configuration with \texttt{anomaly\_fraction} of 0.2 outperforms every configuration with a lower value, consistent with the attack-heavy composition of CIC-DDoS2019, while \texttt{num\_trees} and \texttt{sample\_size} have almost no measurable effect once the anomaly fraction is set appropriately. In feature-vector mode, all nine combinations with anomaly fraction 0.2 tie at a mean F1 of 0.932 and accuracy of 0.873; we selected 200 trees with sample size 128 from this tied set for all live deployments. The best packet-count configuration reaches nearly the same quality (F1 0.929, accuracy 0.868), so the two input modes differ in input representation only; Section~\ref{sec:disc_detection_accuracy} discusses this near-equivalence. The selected configuration is held fixed for every subsequent experiment.

\begin{table}[htbp]
\centering
\caption{Isolation Forest probe results (mean of three runs). The first two rows show the best configuration per input mode; the last two illustrate the spread within packet-count mode and the drop at a lower anomaly fraction.}
\label{tab:if_probe_input_modes}
\begin{tabular}{llccccc}
\hline
\textbf{Result Report} & \textbf{Input mode} & \textbf{Trees} & \textbf{Sample} & \textbf{Anom.\ frac.} & \textbf{F1} & \textbf{Acc.} \\
\hline
best & Feature vector & 200 & 128 & 0.2 & 0.932 & 0.873 \\
best & Packet count   & 200 & 128 & 0.2 & 0.929 & 0.868 \\
worst at 0.2 & Packet count & 50 & 512 & 0.2 & 0.909 & 0.833 \\
best at 0.1 & Packet count & 200 & 512 & 0.1 & 0.830 & 0.710 \\
\hline
\end{tabular}
\end{table}

\subsection{Detection Accuracy Results}

Whereas the previous section used an offline probe to select a detector configuration, this section reports how that selected configuration performs when the NIDS is run live. The live evaluation is carried out on a fixed set of 1{,}533 windows drawn from CIC-DDoS2019 and is repeated across every architecture and detector variant. These windows come from the same dataset that the probe used to rank configurations, and we did not hold out a separate test set; the hyperparameters were therefore both tuned and evaluated on CIC-DDoS2019. The live numbers should accordingly be read as the performance of the deployed system on this dataset, not as an independent test of generalisation to unseen traffic. The difference between the two is that the probe scores a configuration offline over the full dataset, whereas the live run measures the same configuration inside the running NIDS, including its warm-up and online-update behaviour, which is why the live and probe figures do not coincide.

Table~\ref{tab:detection_accuracy} reports the detection quality of all six architecture and detector combinations on the same 1{,}533 windows. The Isolation Forest variants dominate the baseline on every metric except precision, where both are near perfect; the mechanisms behind the differences are analysed in Section~\ref{sec:disc_detection_accuracy}.

\begin{table}[tb]
\centering
\caption{Live detection quality per architecture and detector over 1{,}533 windows.}
\label{tab:detection_accuracy}
\begin{tabular}{llcccc}
\hline
\textbf{Architecture} & \textbf{Detector} & \textbf{Accuracy} & \textbf{Precision} & \textbf{Recall} & \textbf{F1} \\
\hline
Monolithic & Baseline & 0.4047 & 1.000 & 0.3964 & 0.5677 \\
Kafka      & Baseline & 0.4060 & 1.000 & 0.3957 & 0.5670 \\
gRPC       & Baseline & 0.4067 & 1.000 & 0.3972 & 0.5686 \\
Monolithic & Isolation Forest & 0.9335 & 0.9963 & 0.9356 & 0.9650 \\
Kafka      & Isolation Forest & 0.8422 & 0.9992 & 0.8409 & 0.9132 \\
gRPC       & Isolation Forest & 0.9150 & 0.9977 & 0.9151 & 0.9546 \\
\hline
\end{tabular}
\end{table}

\subsection{Software Performance Results}
\label{sec:software_performance_results}

The software performance analysis focuses on two metrics: the per-window processing time and the memory consumption for each software architecture.

Three time metrics are defined for every processed window:

\begin{equation}
\begin{aligned}
\text{Processing Time} &= t_{\text{end}} - t_{\text{start}}, \\
\text{Detector Time} &= t_{\text{detector\_end}} - t_{\text{detector\_start}}, \\
\text{Transport Time} &= \text{Processing Time} - \text{Detector Time},
\end{aligned}
\label{eq:time_metrics}
\end{equation}

where the detector timestamps always bound the detector call itself, and the outer timestamps depend on the architecture: in the monolithic variant, they bound the post-window processing routine, in the Kafka variant, they span from feature message production to result consumption at the engine, and in the gRPC variant, from request emission to response receipt. In the monolithic case, there is no inter-process stage, so the transport time captures only in-process bookkeeping and is near zero.

Two caveats apply. The measured time never includes packet capture or window accumulation, only the work after a window closes. Moreover, the timer starts slightly differently across architectures: the monolithic timer covers the whole post-window routine, including feature extraction, whereas the Kafka and gRPC timers start at message emission, after feature extraction. The reported processing time is therefore a post-window, detector-side latency rather than the full end-to-end latency of the NIDS.

Table~\ref{tab:time_results} reports the average and 99th percentile of each time metric for all six combinations; the 95th percentile and maximum follow the same ordering across variants. The gRPC transport stays below 2 ms on average while the Kafka transport exceeds 27 ms, an order-of-magnitude difference analysed in Section~\ref{chapter:discussion}.

\begin{table}[tb]
\centering
\caption{Per-window time measurements in microseconds (average / 99th percentile).}
\label{tab:time_results}
\setlength{\tabcolsep}{4pt}
\begin{tabular}{llccc}
\hline
\textbf{Architecture} & \textbf{Detector} & \textbf{Processing} & \textbf{Detector} & \textbf{Transport} \\
\hline
Monolithic & Baseline & 352 / 886 & 349 / 878 & 0 / 0 \\
Monolithic & Isolation Forest & 9,901 / 42,164 & 9,898 / 42,161 & 0 / 0 \\
Kafka      & Baseline & 27,922 / 43,472 & 138 / 443 & 27,783 / 43,363 \\
Kafka      & Isolation Forest & 44,418 / 87,870 & 16,231 / 59,266 & 28,186 / 44,484 \\
gRPC       & Baseline & 789 / 2,408 & 143 / 434 & 645 / 2,096 \\
gRPC       & Isolation Forest & 12,244 / 49,554 & 11,368 / 48,886 & 876 / 3,074 \\
\hline
\end{tabular}
\end{table}

\textbf{Memory consumption.}\quad In addition to the time metrics, the memory footprint reported here is the Go runtime heap allocation (\texttt{runtime.MemStats.Alloc}) of the NIDS engine process, sampled once per window. It measures the engine's own footprint only and excludes the separate detector process and, in the Kafka case, the JVM broker, both of which run as separate processes outside the measured one. Table~\ref{tab:memory_results} reports the average and maximum heap allocation of the engine per configuration.

\begin{table}[tb]
\centering
\caption{Engine heap allocation in KB (average / maximum).}
\label{tab:memory_results}
\begin{tabular}{lcc}
\hline
\textbf{Architecture} & \textbf{Baseline} & \textbf{Isolation Forest} \\
\hline
Monolithic & 22,019 / 34,852 & 25,495 / 38,584 \\
Kafka      & 26,595 / 46,209  & 26,475 / 46,389 \\
gRPC       & 21,218 / 34,785  & 21,374 / 34,288 \\
\hline
\end{tabular}
\end{table}

The Kafka engine consistently consumes the most memory, because the in-process Kafka client's buffers and background threads live on the engine heap; the gRPC engine is lightest and the monolithic engine sits between them. The detector affects the footprint less than the transport: switching from baseline to Isolation Forest changes the Kafka and gRPC engines by only a few hundred kilobytes, since their model lives in the separate detector process, while the monolithic engine grows by about 3.5 MB because the model runs inside the measured process. In all cases the footprint stays far below the 16 GB available on the platform, so engine memory does not limit any evaluated configuration.

\section{Discussion}
\label{chapter:discussion}

This section discusses detection accuracy and software performance. Raspberry Pi OS is not a real-time operating system, so individual time measurements vary between runs; we therefore argue from trends across variants rather than from absolute numbers.

\subsection{Detection Accuracy}
\label{sec:disc_detection_accuracy}
When every variant uses the same 75-window warm-up procedure, detection quality on the Raspberry Pi 5 separates first along the detector axis. As Table~\ref{tab:detection_accuracy} shows, the three baseline configurations cluster tightly while all three Isolation Forest configurations sit far above them. Along the transport axis, the picture is more nuanced. For the baseline, the transport has essentially no effect, because the detector receives an identical input and returns an identical verdict whether the window is delivered by an in-process function call, a unary gRPC stub, or a Kafka consumer. For Isolation Forest, the monolithic and gRPC variants are close, but the Kafka variant is about nine percentage points below the monolithic case.

The gap is not a property of the algorithm, since the same Isolation Forest runs in all three deployments. We did not instrument the pipeline to isolate its cause, but our hypothesis is that Kafka's asynchronous delivery changes the order and timing of the online model updates relative to the synchronous in-process and gRPC paths, leaving the model in a slightly different state when a given window is scored; consistently, the Kafka variant also shows the largest transport time. Confirming this would require logging the per-window delivery and update order. The practical implication is that the choice of detector matters far more than the choice of transport: a synchronous transport preserves in-process detection quality, while a heavily decoupled transport can introduce a measurable detection penalty that must be weighed against the durability and decoupling it provides.

The baseline detector has a precision of 1 but low recall, indicating that it produces no false positives but misses the labelled attack windows that contain low-volume malicious traffic. Isolation Forest achieves a precision on par with the baseline (ranging from 0.996 to 0.999) while its recall climbs, which also increases the F1. This reflects that Isolation Forest widens the alarm region and suggests that it has more potential in a DDoS detection scenario, where subtle and low-volume attack patterns should be discovered and flagged early.

The near-equivalence of the feature-vector and packet-count inputs follows from the nature of CIC-DDoS2019~\cite{cicddos2019_paper}, which mostly contains high-volume flooding attacks: when the attack signal is a sharp volume increase, the packet count alone is a near-sufficient indicator, so the additional flow features add little separation at the window level. This does not make the richer vector redundant in general; feature selection is left as future work.
\subsection{Software Performance Comparison}

Regarding transport time, gRPC has a lower transport time compared to Kafka. The average transport time of gRPC is less than 2 milliseconds; however, the average transport time of Kafka is more than \SI{27}{\milli\second}. This indicates that the cost of switching from an in-process call to a network-based microservice is not uniform across transports: the Kafka path imposes more than an order of magnitude more transport overhead than the gRPC path on the same Raspberry Pi 5 host.

Since both microservice variants communicate over loopback on the same host, the difference reflects protocol design rather than network cost. The Kafka budget pays for producer batching, the broker-side log append and acknowledgement, and the consumer fetch cycle; the broker's \texttt{fetch.max.wait.ms} alone introduces a default delay of this order. The measured \SI{27}{\milli\second} is thus the expected cost of the durability, decoupling, and replay properties that Kafka provides beyond gRPC, whereas a synchronous unary gRPC call avoids these mechanisms and stays in the low milliseconds.

\subsection{Switching from Baseline to Isolation Forest}

\begin{figure}[H]
    \centering
    \includegraphics[width=0.65\linewidth]{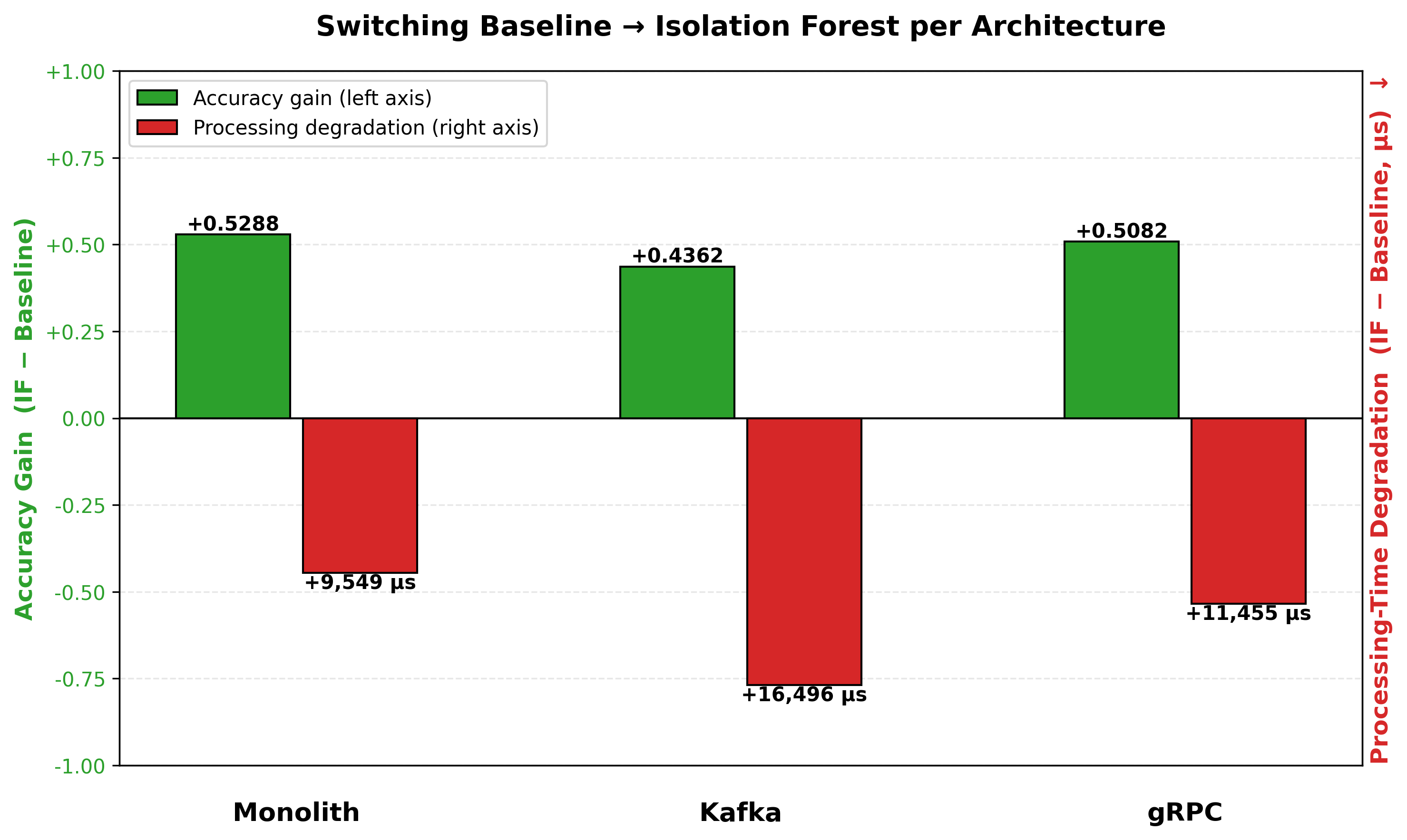}
    \caption{Accuracy gain and processing time degradation overview}
    \label{fig:comparison}
\end{figure}

Figure~\ref{fig:comparison} shows the improvements and degradation when switching from the baseline model to the Isolation Forest. The switch produces a much higher accuracy across all three software architectures, but it also degrades the processing time in every case. The microservice architecture with Kafka shows the largest degradation, while the monolithic architecture shows the smallest.

This shows a trade-off between accuracy and processing time. For practitioners who require a microservice architecture, for instance, to decouple the detection pipeline in cloud-based deployments or a multi-process structure, the gRPC variant emerges as the strongest candidate, since it preserves a comparable accuracy gain while incurring a smaller processing-time degradation than the Kafka variant. 

One thing that should be read with caution is that, based on the dataset used in this paper, the monolithic architecture appears to be the most favourable choice, as it combines the accuracy gain with the smallest processing-time degradation. However, CIC-DDoS2019 mainly contains high-volume attack traffic but might not necessarily exercise the monolithic architecture under conditions in which a single process becomes a bottleneck. It is therefore possible that the monolithic advantage observed here partly reflects the characteristics of the dataset rather than a general property of the architecture. As a result, a systematic study of the conditions under which the monolithic architecture begins to degrade, and of how the microservice variants compare under heavier or more diverse workloads, is left as future work.

\subsection{Effectiveness of Sequential Replayer}
\label{sec:limitation_replayer}

Wickman and Rygaard~\cite{wickmannidschalmers} noted that an in-process replayer risks becoming the bottleneck when the required inter-packet delay is shorter than its processing time. In our setup the replayer runs as an independent program on the attacker device, so its speed bounds the attack, not the NIDS. Replaying without artificial inter-packet delays reached an average of 3 Gbps, roughly three times the peak attack rate captured in CIC-DDoS2019, so the replayer does not constrain the evaluation.

\subsection{Limitations}

\subsubsection{Feature Selection}

Only two input configurations were evaluated: a single packet count and the full 80-dimensional vector, the two extremes of the feature space. Isolation Forest is sensitive to irrelevant or redundant dimensions~\cite{isolation_forest_sensitive}, so a well-chosen subset might improve detection quality, and a smaller vector would simultaneously reduce scoring, serialisation, and transport costs in the microservice variants. The results should therefore be read as comparing a minimal and a complete representation, not as the best achievable trade-off.

\subsubsection{Validity Threats}

Several validity threats qualify these results. Internally, the experiments ran on a non-isolated Linux system: background tasks, dynamic CPU frequency, and the JVM garbage collection of the Kafka broker add noise unrelated to the architectures, and the window labels inherit any inaccuracies of the CIC-DDoS2019 ground truth, which was itself generated with CICFlowMeter. Externally, the Raspberry Pi 5 is a single hardware configuration and CIC-DDoS2019 is a synthetic dataset from 2019, so the findings indicate behaviour in an embedded-like environment rather than values that transfer to arbitrary hardware or present-day traffic; the hyperparameters were moreover tuned and evaluated on the same dataset. Regarding constructs, window-level labelling can hide short bursts and does not guarantee that a detector fires for the intended reason, the monolithic transport time is structurally zero rather than an empirical result, and the conclusions apply to these two concrete detector implementations rather than to algorithm classes. Finally, mitigation remains per-IP with a count-based per-IP detector, which is weaker against attacks distributed across many low-rate sources.

\subsection{Future Work}

Future work includes three directions. Feature selection could identify an informative subset of the 80 features and quantify its joint effect on detection quality and on per-architecture serialisation and transport cost. Sequence models such as LSTM report strong detection performance~\cite{nids_lstm,cicddos2019_paper,xbpf_xdp_smartx} and could replace Isolation Forest, at a resource cost that must be evaluated on the same constrained hardware. The eBPF/XDP mitigation could move from permanent to time-bounded blocks, so that a falsely flagged address regains access, eventually allowing the rate limiter to be retired in favour of the detector alone.

\subsection{Ethics}

This work is defensive in purpose. It uses only the public, synthetically generated CIC-DDoS2019 dataset, which contains no personal data, and no attempt was made to identify any host or individual represented in it. All attack replay was confined to an isolated testbed against our own device, and no traffic was directed at third parties or the public internet. The released artefacts, in particular GoFlowMeter, are general-purpose flow-analysis tools rather than attack software, which limits their potential for misuse.

\section{Conclusion}
\label{chapter:conclusion}
This research addressed DDoS attack mitigation through two research questions (Section~\ref{sec:research_questions}): how to improve detection effectiveness using modern machine learning and fast packet filtering, and how system architecture affects performance. 

We extended an existing NIDS with eBPF/XDP integration, the GoFlowMeter feature extractor, a standalone attacker program, a fine-tuned Isolation Forest model, and a comparison of three deployment architectures (Section~\ref{sec:contribution}).

Answering the first research question, the fine-tuned Isolation Forest with the 80-dimensional feature-vector input (200 trees, sample size 128, anomaly fraction 0.2) reached a live accuracy of 0.9335 and an F1 of 0.9650 in the monolithic deployment (Table~\ref{tab:detection_accuracy}), clearly improving on the statistical baseline, whose recall suffers on low-volume attack windows. The feature-vector input also outperformed the packet-count input, although only narrowly on this flood-dominated dataset.

Answering the second research question, the architecture comparison showed that the gRPC-based microservice deployment preserves near-monolithic detection quality with an average transport time below 2 ms, whereas the Kafka-based deployment adds over 27 ms of transport time and a measurable detection penalty; the monolithic deployment remains the most favourable on this dataset, with the caveat that the workload may not expose single-process bottlenecks.

The multi-feature detection provided by GoFlowMeter resolves the single-threshold adaptability problem left open by Wickman and Rygaard~\cite{wickmannidschalmers}, although detecting genuinely slow-growing attacks remains to be demonstrated on datasets that contain more of them.

Future work includes feature selection, evaluating sequence models such as LSTM on the same constrained hardware, and time-bounded blocking in the eBPF/XDP mitigation.

\begin{credits}
\subsubsection{\ackname} This work was carried out in collaboration with Ericsson's Radio and Transport Engineering division, and is based on the master's thesis of the first two authors at Chalmers University of Technology and the University of Gothenburg.

\subsubsection{\discintname} Shiqi Wu, Oleksii Koshovyi, and Georgios Pseiridis Pseiras are employed by Ericsson AB. The authors have no other competing interests to declare that are relevant to the content of this article.
\end{credits}
%
%
%
\bibliographystyle{splncs04}
\bibliography{references}

@misc{xbpf_xdp_smartx,
      title={Smart{X} {I}ntelligent {Sec}: A Security Framework Based on Machine Learning and {eBPF/XDP}}, 
      author={Talaya Farasat and JongWon Kim and Joachim Posegga},
      year={2024},
      eprint={2410.20244},
      archivePrefix={arXiv},
      primaryClass={cs.CR},
      url={https://arxiv.org/abs/2410.20244}, 
}

@conference{CIC-IDS2017,
author={Iman Sharafaldin and Arash {Habibi Lashkari} and Ali A. Ghorbani},
title={Toward Generating a New Intrusion Detection Dataset and Intrusion Traffic Characterization},
booktitle={Proceedings of the 4th International Conference on Information Systems Security and Privacy - ICISSP},
year={2018},
pages={108-116},
publisher={SciTePress},
organization={INSTICC},
doi={10.5220/0006639801080116},
isbn={978-989-758-282-0},
issn={2184-4356},
}

@mastersthesis{wickmannidschalmers,
  author  = "Albert Wickman and Daniel Rygaard",
  title   = "Scalable Anomaly-Based Network Intrusion Detection Using a Statistical Model and Data Sketches",
  school  = "Chalmers University of Technology",
  year    = "2025",
  url = "https://hdl.handle.net/20.500.12380/310882"
}

@INPROCEEDINGS{cicddos2019_paper,
  author={Sharafaldin, Iman and Lashkari, Arash Habibi and Hakak, Saqib and Ghorbani, Ali A.},
  booktitle={2019 International Carnahan Conference on Security Technology (ICCST)}, 
  title={Developing Realistic Distributed Denial of Service ({DDoS}) Attack Dataset and Taxonomy}, 
  year={2019},
  volume={},
  number={},
  pages={1-8},
  doi={10.1109/CCST.2019.8888419}}

@misc{caida2007,
  author       = {Vaishali Shirsath},
  title        = {CAIDA UCSD DDoS 2007 Attack Dataset},
  year         = {2023},
  howpublished = {IEEE Dataport},
  doi          = {10.21227/dvp9-s124},
  url          = {https://dx.doi.org/10.21227/dvp9-s124}
}

@article{hids_vs_nids,
author = {Bridges, Robert A. and Glass-Vanderlan, Tarrah R. and Iannacone, Michael D. and Vincent, Maria S. and Chen, Qian (Guenevere)},
title = {A Survey of Intrusion Detection Systems Leveraging Host Data},
year = {2019},
issue_date = {November 2020},
publisher = {Association for Computing Machinery},
address = {New York, NY, USA},
volume = {52},
number = {6},
issn = {0360-0300},
url = {https://doi.org/10.1145/3344382},
doi = {10.1145/3344382},
journal = {ACM Comput. Surv.},
month = nov,
articleno = {128},
numpages = {35}
}

@INPROCEEDINGS{isolation_forest_4781136,
  author={Liu, Fei Tony and Ting, Kai Ming and Zhou, Zhi-Hua},
  booktitle={2008 Eighth IEEE International Conference on Data Mining}, 
  title={Isolation Forest}, 
  year={2008},
  volume={},
  number={},
  pages={413-422},
  doi={10.1109/ICDM.2008.17}}

@inproceedings{cicflowmeter1,
  title={Characterization of {T}or traffic using time based features},
  author={Lashkari, Arash Habibi and Gil, Gerard Draper and Mamun, Mohammad Saiful Islam and Ghorbani, Ali A},
  booktitle={International conference on information systems security and privacy},
  volume={2},
  pages={253--262},
  year={2017},
  organization={SciTePress}
}

@conference{cicflowmeter2,
author={Gerard Draper{-}Gil and Arash Habibi Lashkari and Mohammad Saiful Islam Mamun and Ali {A. Ghorbani}},
title={Characterization of Encrypted and {VPN} Traffic using Time-related Features},
booktitle={Proceedings of the 2nd International Conference on Information Systems Security and Privacy - ICISSP},
year={2016},
pages={407-414},
publisher={SciTePress},
organization={INSTICC},
doi={10.5220/0005740704070414},
isbn={978-989-758-167-0},
issn={2184-4356},
}

@misc{goflowmeter,
  author = {Shiqi Wu and Oleksii Koshovyi},
  title = {{Goflowmeter: A High-Performance Network Flow Feature Extractor in Go}},
  year = {2024},
  publisher = {GitHub},
  journal = {GitHub repository},
  howpublished = {\url{https://anonymous.4open.science/r/goflowmeter-E465/README.md}},
  note = {Accessed: 2026-03-11}
}

@INPROCEEDINGS{monolithic_definition,
  author={Kuryazov, Dilshodbek and Jabborov, Dilshod and Khujamuratov, Bekmurod},
  booktitle={2020 IEEE 14th International Conference on Application of Information and Communication Technologies (AICT)}, 
  title={Towards Decomposing Monolithic Applications into Microservices}, 
  year={2020},
  volume={},
  number={},
  pages={1-4},
  doi={10.1109/AICT50176.2020.9368571}}

@misc{cloudflare_aisuru_botnet_report,
  author       = {{Cloudflare}},
  title        = {Aisuru botnet: Early October attacks escalate into record-setting {DDoS} activity},
  year         = {2025},
  month        = {12},
  url          = {https://www.cloudflare.com/threat-intelligence/research/report/aisuru-botnet/},
  note         = {Accessed: 2026-03-18}
}

@article{ferreira2026botnet,
  author       = {Ferreira, Bruno},
  title        = {Botnet smashes {DDoS} traffic record, equivalent to streaming 2.2 million Netflix 4K movies at once — 31.4 Tb/s attack was large enough to take entire countries offline},
  journal      = {Tom's Hardware},
  year         = {2026},
  month        = {01},
  url          = {https://www.tomshardware.com/service-providers/network-providers/botnet-smashes-ddos-traffic-record-at-31-4-tb-s-equivalent-to-streaming-2-2-million-netflix-4k-movies-at-once-attack-was-large-enough-to-take-entire-countries-offline},
  note         = {Accessed: 2026-03-18}
}

@article{nids_cnn,
title = {A new {DDoS} attacks intrusion detection model based on deep learning for cybersecurity},
journal = {Computers \& Security},
volume = {118},
pages = {102748},
year = {2022},
issn = {0167-4048},
doi = {10.1016/j.cose.2022.102748},
author = {Devrim Akgun and Selman Hizal and Unal Cavusoglu}
}

@article{nids_lstm,
author = {Shurman, Mohammad and Khrais, Rami and Yateem, A.Rahman},
year = {2020},
month = {07},
pages = {655-661},
title = {DoS and DDoS Attack Detection Using Deep Learning and IDS},
volume = {17},
journal = {International Arab Journal of Information Technology},
doi = {10.34028/iajit/17/4A/10}
}

@INPROCEEDINGS{nids_cnn_Shaaban,
  author={Shaaban, Ahmed Ramzy and Abd-Elwanis, Essam and Hussein, Mohamed},
  booktitle={2019 Ninth International Conference on Intelligent Computing and Information Systems (ICICIS)}, 
  title={DDoS attack detection and classification via Convolutional Neural Network (CNN)}, 
  year={2019},
  volume={},
  number={},
  pages={233-238},
  doi={10.1109/ICICIS46948.2019.9014826}}

@INPROCEEDINGS{nids_lstm_djama,
  author={Djama, Asma and Maazouz, Mohamed and Kheddar, Hamza and Rahim, Messaoud and Bengherbia, Billel},
  booktitle={2025 International Conference on Intelligent Computer Systems, Data Science and Applications (IC2SDA)}, 
  title={High Performance Detection of DDoS Attacks with Hybrid CNN-LSTM Architecture}, 
  year={2025},
  volume={},
  number={},
  pages={1-6},
  doi={10.1109/IC2SDA68097.2025.11331388}}

@INPROCEEDINGS{nids_random_forest_bhawsar,
  author={Bhawsar, Pranav and Kumar, Abhinav and Kumar, Sunny and Gupta, Vikas and Rana, Rahul and Mehta, Prachi and Mishra, Zeesha},
  booktitle={2025 13th International Conference on Intelligent Systems and Embedded Design (ISED)}, 
  title={Machine Learning-Based Detection of DDoS Attacks: An Evaluation of SVM and Random Forest Approaches}, 
  year={2025},
  volume={},
  number={},
  pages={61-67},
  doi={10.1109/ISED67359.2025.11405254}}

@INPROCEEDINGS{ghani2025,
  author={Ghani, Elang Prasakti and Sofwan, Aghus and Somantri, Maman},
  booktitle={2025 International Conference on Smart Computing, IoT and Machine Learning (SIML)}, 
  title={AI-Driven Network Security: Detecting and Mitigating DDoS, Malware, and Backdoor Attacks with Isolation and Random Forest Algorithm}, 
  year={2025},
  volume={},
  number={},
  pages={1-6},
  doi={10.1109/SIML65326.2025.11080951}}

@inproceedings{vinothina2025,
  title={An efficient {DDoS} attack detection scheme using autoencoder based on feature selection approaches},
  author={Vinothina, V and Prakash, VS and Revathi, K and Kumar, C Sathish and Karthikeyen, N and Prathap, G},
  booktitle={2025 Third International Conference on Augmented Intelligence and Sustainable Systems (ICAISS)},
  pages={372--377},
  year={2025},
  organization={IEEE}
}

@inproceedings{xdp_10.1145/3281411.3281443,
author = {H\o{}iland-J\o{}rgensen, Toke and Brouer, Jesper Dangaard and Borkmann, Daniel and Fastabend, John and Herbert, Tom and Ahern, David and Miller, David},
title = {The e{X}press data path: fast programmable packet processing in the operating system kernel},
year = {2018},
isbn = {9781450360807},
publisher = {Association for Computing Machinery},
address = {New York, NY, USA},
doi = {10.1145/3281411.3281443},
booktitle = {Proceedings of the 14th International Conference on Emerging Networking EXperiments and Technologies (CoNEXT)},
pages = {54–66},
numpages = {13},
location = {Heraklion, Greece},
series = {CoNEXT '18}
}

@misc{dataset-xdp-farasat2024,
      title={Advancing Network Security: A Comprehensive Testbed and Dataset for Machine Learning-Based Intrusion Detection}, 
      author={Talaya Farasat and JongWon Kim and Joachim Posegga},
      year={2024},
      eprint={2410.18332},
      archivePrefix={arXiv},
      primaryClass={cs.CR},
      url={https://arxiv.org/abs/2410.18332}, 
}

@article{Goldschmidt_2025,
   title={Network intrusion datasets: A survey, limitations, and recommendations},
   volume={156},
   ISSN={0167-4048},
   url={http://dx.doi.org/10.1016/j.cose.2025.104510},
   DOI={10.1016/j.cose.2025.104510},
   journal={Computers \& Security},
   publisher={Elsevier BV},
   author={Goldschmidt, Patrik and Chudá, Daniela},
   year={2025},
   month=sep, pages={104510} }

@misc{raspberry_pi_spec,
  author       = {{Raspberry Pi}},
  title        = {Raspberry Pi 5},
  year         = {2026},
  month        = {05},
  url          = {https://www.raspberrypi.com/products/raspberry-pi-5/},
  note         = {Accessed: 2026-05-05}
}

@misc{apache_kafka,
  author       = {{Apache Software Foundation}},
  title        = {Apache Kafka},
  year         = {2026},
  month        = {05},
  url          = {https://kafka.apache.org/},
  note         = {Accessed: 2026-05-05}
}

@misc{grpc_website,
  author       = {{gRPC Authors}},
  title        = {Introduction to gRPC},
  year         = {2026},
  month        = {05},
  url          = {https://grpc.io/docs/what-is-grpc/introduction/},
  note         = {Accessed: 2026-05-05}
}

@INPROCEEDINGS{protocol_buffer,
  author={Popić, Srđan and Pezer, Dražen and Mrazovac, Bojan and Teslić, Nikola},
  booktitle={2016 International Conference on Smart Systems and Technologies (SST)}, 
  title={Performance evaluation of using Protocol Buffers in the Internet of Things communication}, 
  year={2016},
  volume={},
  number={},
  pages={261-265},
  doi={10.1109/SST.2016.7765670}}

@Article{isolation_forest_sensitive,
AUTHOR = {Heigl, Michael and Anand, Kumar Ashutosh and Urmann, Andreas and Fiala, Dalibor and Schramm, Martin and Hable, Robert},
TITLE = {On the Improvement of the Isolation Forest Algorithm for Outlier Detection with Streaming Data},
JOURNAL = {Electronics},
VOLUME = {10},
YEAR = {2021},
NUMBER = {13},
ARTICLE-NUMBER = {1534},
URL = {https://www.mdpi.com/2079-9292/10/13/1534},
ISSN = {2079-9292},
DOI = {10.3390/electronics10131534}
}

@INPROCEEDINGS{kafka_motivation1,
  author={Gogineni, Vineeth},
  booktitle={2025 IEEE 6th International Seminar on Artificial Intelligence, Networking and Information Technology (AINIT)}, 
  title={Apache Kafka for Low Latency AI-Enhanced Data Analytics for Fraud Detection System}, 
  year={2025},
  volume={},
  number={},
  pages={1720-1724},
  doi={10.1109/AINIT65432.2025.11035547}}

@INPROCEEDINGS{grpc_motivation1,
  author={Nimpattanavong, Chollakorn and Khan, Ibrahim and Van Nguyen, Thai and Thawonmas, Ruck and Choensawat, Worawat and Sookhanaphibarn, Kingkarn},
  booktitle={2023 8th International Conference on Business and Industrial Research (ICBIR)}, 
  title={Improving Data Transfer Efficiency for AIs in the DareFightingICE using gRPC}, 
  year={2023},
  volume={},
  number={},
  pages={286-290},
  doi={10.1109/ICBIR57571.2023.10147629}}

@article{tebbaa2025mitigating,
  title={Mitigating {DDoS} attacks in software-defined networks: a systematic literature review of machine learning and deep learning approaches},
  author={Tebbaa, Kaoutar and Chakir, Oumaima and Maleh, Yassine and Bela{\"\i}ssaoui, Mustapha},
  journal={Iranian Journal of Computer Science},
  year={2025},
  publisher={Springer Nature},
  doi={10.1007/s42044-025-00386-x},
  url={https://doi.org/10.1007/s42044-025-00386-x}
}

@article{perfromance_impact,
author = {Reis, Manuel and Serodio, Carlos},
year = {2025},
month = {04},
pages = {179},
title = {Edge {AI} for Real-Time Anomaly Detection in Smart Homes},
volume = {17},
journal = {Future Internet},
doi = {10.3390/fi17040179}
}

%




\end{document}